\documentclass[9pt]{article}

\usepackage[export]{adjustbox}
\usepackage{amsthm}
\usepackage{nccmath}
\usepackage{balance}
\usepackage{etoolbox}

\usepackage[utf8]{inputenc}
\usepackage{flushend}
\usepackage[section]{placeins}
\usepackage[bf, footnotesize]{caption}
\usepackage{graphics}
\usepackage{CJKutf8}

\usepackage{amssymb}
\usepackage{mathtools, cuted}
\usepackage{widetext}

\usepackage{lineno}

\usepackage{multicol}
\usepackage{booktabs}
\usepackage{amsmath}
\usepackage{widetext}
\usepackage{flushend}
\usepackage[explicit]{titlesec}
\usepackage{authblk}
\titleformat{\section}
  {\normalfont}{\thesection}{1em}{\MakeUppercase{#1}}
  \titleformat{\subsection}
  {\it}{\thesection}{1em}{{#1}}
  
\begin{document}
\small

\title{Virtualization of Classical Reality: Limits and Possibilities in Physical Simulation}
\date{}


\author[1,2]{Francesco Sisini}
\affil[1]{
  VERGANI NAVARRA HIGHER EDUCATION INSTITUTE
  - Department of Mathematics and Physics,
  Via Sogari 3 - 44121 Ferrara, Italy
}
\affil[2]{Department of Applied Research Tekamed Ltd. Via Bellaria 6, 44121 Ferrara ,Italy}



\begin{center}

\end{center}
\maketitle

\pagebreak


\section*{Keywords}
Simulation, virtual reality, inertial observer, Oculus, physical computation.

\section*{Abstract}
This study delves into the virtualization of classical reality, aiming to establish a clear framework that elucidates the limits governing this process and provides definitive answers regarding virtual reality. The primary questions explored are whether an observer's senses can be stimulated through appropriate equipment to perceive a different reality from their current environment and whether it is feasible to simulate a reality where the laws of physics do not hold.

With virtual and augmented reality increasingly employed in various fields such as production, maintenance, education, and entertainment, it becomes crucial to provide well-founded responses to these inquiries. The existence of limitations on the precision and achievability of virtual reality is of paramount importance in creating realistic environments for didactic, recreational, or occupational purposes. Conversely, addressing the second question sheds light on the pivotal role of physics and scientific rigor, even within virtual contexts.

The study's objective is to present a theoretical framework that enables precise and contextually relevant responses. It is divided into three sections: the Methods section elucidates the nature of computers and their capacity to create a perceived virtual reality. The Results section introduces the foundational theoretical framework, focusing on the concepts of observable simulation and interactive simulation, and examining their distinctions. Finally, the Discussion section builds upon the theoretical foundation to answer the posed questions and provide comprehensive insights.

This study contributes to a better understanding of the boundaries and possibilities of virtual reality, ensuring that conjecture is avoided and concrete answers are provided. By exploring the potential for creating convincing simulations and examining the adherence to physical laws in virtual environments, this research provides essential knowledge for the development and utilization of virtual reality in diverse applications.

\section*{Introduction}

One possible definition of virtual reality, obtained by directly asking ChatGPT, is that "virtual reality is a technology that creates a computer-simulated environment that can be explored and interacted with by users. Using devices like VR headsets and controllers, users can immerse themselves in a virtual environment that can be both realistic and fantastical. Virtual reality provides an engaging sensory experience, which can include 360-degree panoramic views, three-dimensional audio, and intuitive interactions. The goal of virtual reality is to provide a simulated experience that makes users feel as if they are truly present and engaged in the virtual environment, creating a sense of 'presence' in a digitally generated reality."

For years, virtual reality and immersive experiences have reached a high level of quality, prompting some thinkers to question the possibility of confusing human beings about the reality of the experience \cite{bostrom}. A comprehensive analysis of this topic certainly involves psychological and technological aspects, as well as physical nature. Several authors, from Feynman to Campbell, have addressed related topics, focusing primarily on whether a classical or quantum computer can provide an accurate representation of a physical process \cite{Feynman,campbell}.

Feynman's analysis starts with the question, "Can physics be simulated by a universal computer?" In his writing, Feynman briefly refers to the concept of classical simulation as solving the differential equations that describe a given physical system. He then clarifies that the truly open question is whether the state of a computer can evolve over time exactly like the (quantum) system it is simulating. Campbell also specifically addresses the question of whether tests can be formulated to distinguish reality from simulation. He proposes conceptual quantum experiments designed to test the "simulation theory" \cite{bostrom,whitworth,campbell2}, aiming to identify any inconsistencies or behaviors that might suggest a simulated nature of reality.

The research conducted in this direction, starting from the 1980s with Feynman to the present day, has produced very interesting results regarding the simulation of quantum systems, partly assuming the answer to simple classical problems about the limits that constrain the representations of reality produced by computers for more common experiences largely dominated by classical physics. In practice, based on the fact that for any arbitrarily complex physical system and a set of initial conditions, a solution to its equations can always be calculated with arbitrary precision, it has always been assumed that a universal computer can simulate a classical system with arbitrary precision. This idea is also extended to the context of virtualizing reality because the problems highlighted by Campbell for quantum systems do not arise for classical ones.

The absence of limits to simulating finite classical systems may lead one to think that there are no limits to producing virtual realities for classical systems. Since virtual reality is applied in an increasing number of contexts, it is important to clarify this point and avoid leaving room for conjecture where certain answers can be provided.

To this end, this work aims to outline a clear framework of the limits governing the virtualization of classical reality, allowing us to answer key questions about virtual reality. In practice, we ask ourselves if, after wearing appropriate equipment, our senses can be stimulated in a way that leads us to perceive a reality different from the one we are living in. For example, a properly equipped person placed in an empty room could believe they are in a completely different environment, such as a party or even a tropical jungle. In summary, we pose two specific questions:
\begin{itemize}
\item Is it possible to simulate reality in a way that would lead a (rational) observer to agree that reality and simulation are indistinguishable?
\item Is it possible to simulate a reality where the laws of physics do not apply?
\end{itemize}
Since virtual reality and augmented reality are now widely employed in production processes, maintenance, education, and entertainment, the importance of providing a well-founded answer to the first question is evident. The existence of a limit to the precision or feasibility of virtual reality is a crucial element in all applications where a realistic environment is desired for educational, recreational, or work purposes. The second question investigates another aspect, sometimes overlooked and often assumed as obvious, namely, whether it is possible to violate the laws of physics and experience magical phenomena in a virtual environment. The answer to this question clarifies the crucial role of physics and scientific rigor in any context, even a virtual one.

The purpose of this work is to present a theoretical framework that allows for precise and contextual answers to these two questions.

This work is divided into three sections. The Methods section describes what a computer is and how this tool can create a virtual reality perceived by an observer. The Results section presents the theoretical framework underlying the subsequent reasoning, which is developed in the third section dedicated to the discussion. In particular, the concepts of "observable simulation" and "interactive simulation" are presented and their difference is discussed.

\section{Methods}

To define the theoretical framework for addressing the two questions posed in the introduction, it is necessary to define the conceptual tools on which this framework is based. These tools include:

\begin{itemize}
\item A computerized system for virtual reality production.
\item An observer capable of observing and interacting with both reality and virtual reality.
\item A definition of the simulation process of a physical system.
\end{itemize}

\subsection{Computerized System for Virtual Reality Production}

To produce virtual reality, it is necessary to execute software on a computer and have a human-computer interface through appropriate equipment. To provide a physical description of virtual reality, it is necessary to observe computers themselves in terms of physical systems.

The computer is an artificial physical system $\mathfrak{C}$ whose components follow the laws of physics, and like any other physical system, they can be described using generalized coordinates, constraints, kinetic energy, and potential energy \cite{Dirritch, bennet, Horsman}. Therefore, we can think of the computer as consisting of hardware and software.

\subsubsection{Hardware}

The hardware component is described by a certain number of internal coordinates (referred to as internal because they do not represent a position in three-dimensional space but an internal dimension such as spin) that can in principle be observed using dedicated instruments, such as a multimeter.

Each of these coordinates represents the state of a memory register, which, for simplicity, we can think of as a capacitor that can store a charge quantity $Q_0$ or $Q_1$, corresponding to the storage of the values 0 or 1. For convenience, we group these registers into more complex registers that can take a decimal value with arbitrary finite precision.

Many of these registers are necessary for the functioning of the system; we will refer to them as system registers and will not consider them further. The remaining registers are reserved for the software application for creating virtual reality and are referred to as application registers. For convenience, we consider these registers arranged as a two-dimensional matrix of $N$ rows and $M$ columns, denoted as $\mathcal{C}_{N\times M}$.

\subsubsection{Software}

The software needs to be divided into system software and application software. System software, like system registers, is necessary for the functioning of the computer itself and is not considered in this analysis, which focuses on application software dedicated to creating virtual reality.

The application software represents the initial conditions of the application registers. In fact, at a given instant $t_0$ when a computation begins to create virtual reality, the code loaded into memory is an distribution of electric charge $\rho(x,y,z,t_0)$ represented by the matrix $\mathcal{Q}$ whose elements can only take on the values $Q_0$ or $Q_1$.

\subsubsection{Equipment}

To create a virtual reality system, it is necessary to involve at least one of the observer's senses, such as vision. Additionally, the observer must have the ability to interact to some extent with the virtualized reality. Therefore, we consider the minimum equipment consisting of a vision system called "oculus" and a handheld device called a "controller" that allows some level of interaction with what is displayed by the oculus.

\paragraph{Oculus}
We can think of Oculus as a physical system $\mathfrak{V}$ consisting of a screen that produces images of $R\times C$ pixels, where $R$ and $C$ represent the number of rows and columns of the screen, respectively.

The combination of Oculus and computer, from a physical point of view, constitutes a system $\mathfrak{C}\cup \mathfrak{V}$ with the software representing the initial conditions.

The pixels of Oculus are internal coordinates of the $\mathfrak{V}$ system. Each pixel can take a certain value, which is translated into an observable color. When Oculus is connected to the computer, the color of each pixel is associated with a specific memory register of the computer, so each pixel is bound to a degree of freedom of the computer.

The state of the $\mathfrak{V}$ system can be represented by a matrix of values $\mathcal{V}_{R\times C}(t)$, where each element of the matrix represents the value associated with the corresponding pixel. There is a relationship between the elements of $\mathcal{V}$ and $\mathcal{C}$, given by $\mathcal{V}{r,c}=\mathcal{C}_{r,c}$.

During computation, the distribution of charge among the various application registers changes over time due to the computation itself, and as a result, the color presented by the Oculus pixels also changes.
Each pixel follows a precise temporal law defined by the equation describing the dependence of the matrix $\mathcal{C}$ on time. This law is encoded in the software executed by the computer:
\begin{equation}
\mathcal{V}{r,c}(t)=\mathcal{C}{r,c}(t)=f_i(x_1,...,x_{n})
\end{equation}
where $x_1,...,x_n$ are the internal degrees of freedom of the software, and the index $i$ is calculated as $i=(r-1)\times C +c$, and $n=N\times M$.

\paragraph{Controller}
A controller is a handheld object with six or more degrees of freedom described by as many coordinates, where six of them specify its position and orientation in space, and the remaining ones may represent the state and/or activation/deactivation of switches and buttons, which are considered internal degrees of freedom.
The state of the controller can be represented by a column matrix $\mathcal{X}$ whose values are mapped to the matrix $\mathcal{C}$ representing the state of the computer. In this way, the physical actions performed on the controller modify the dynamic evolution of the computation and can also modify what is displayed by the Oculus.

We have provided a rigorous definition of the computerized system for virtual reality production. To complete the theoretical framework, we need to introduce the concept of an inertial observer who is capable of observing both the computerized system as a physical system and the virtual reality produced by that system. The observer will serve us to give an objective definition of virtual reality.
\subsection{Inertial Observer}
\begin{figure}
\centering
\includegraphics[width=0.8\textwidth]{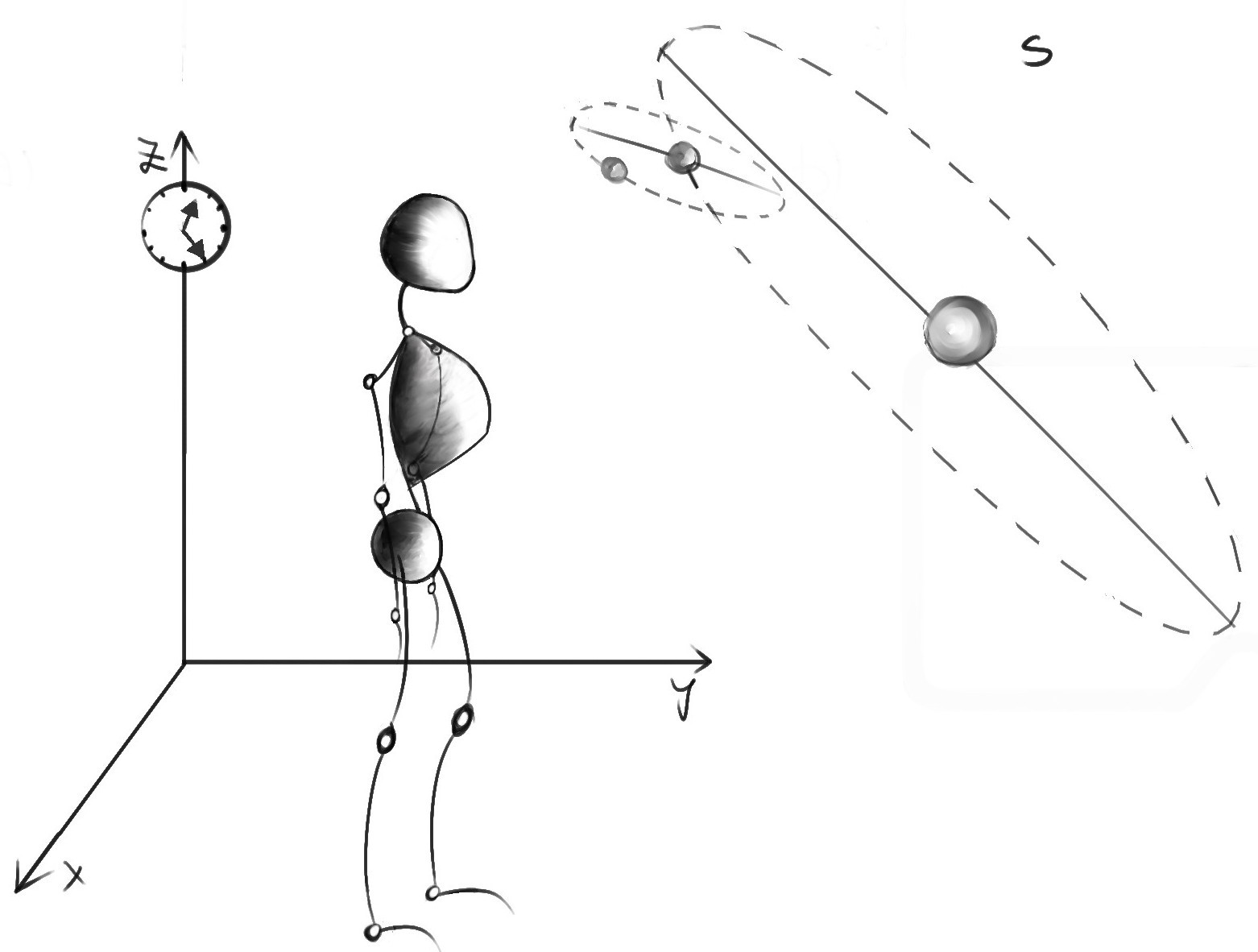}
\caption{The inertial observer observes a physical system $\mathfrak{C}$ consisting of three celestial bodies (Sun, Earth, and Moon) in orbit according to the law of universal gravitation.}
\label{fig:io1}
\end{figure}

Let's introduce a well-known concept in classical physics, namely the inertial observer, who, equipped with a clock and a reference system, observes a certain physical system $\mathfrak{S}$ (see Figure \ref{fig:io1}).

The observer can measure the position of each of the $n$ coordinates $q_i$ that describe the system. In the most general case, the coordinates of the system evolve in time according to certain time laws that the observer can track. Furthermore, suppose that the observer is aware of the known laws of physics.


Additionally, let's consider that the observed motions are sufficiently slow to be described by assuming instantaneous interaction between the components of the system $\mathfrak{C}$, allowing for a non-relativistic description of the dynamics.

\begin{figure}
\centering
\includegraphics[width=0.8\textwidth]{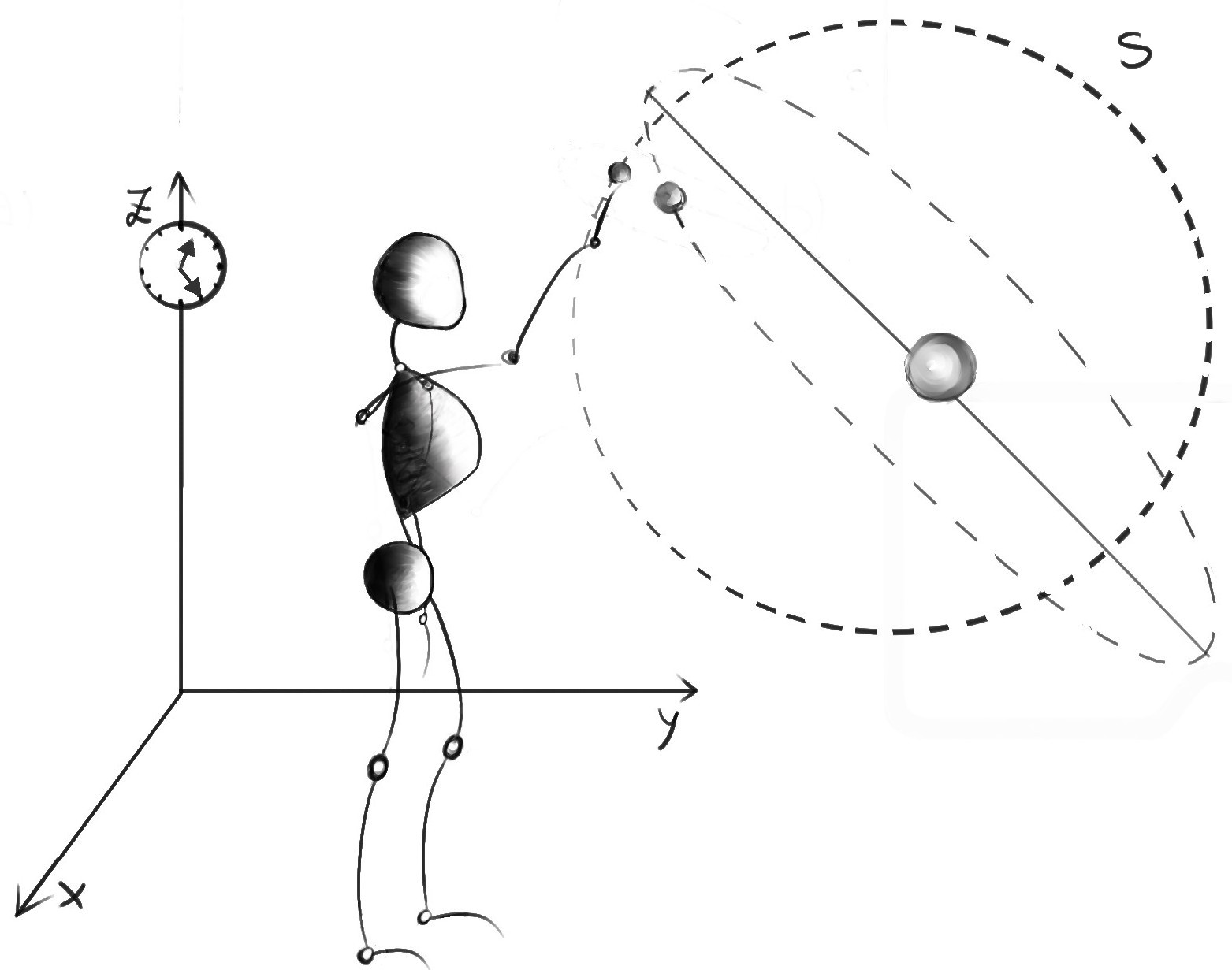}
\caption{The (interacting) inertial observer applies a force $\vec{F}_e$ to the physical system $\mathfrak{C}$, resulting in a new orbit for the Moon.}
\label{fig:io2}
\end{figure}

If the observer is informed about the initial conditions $q_i(t_0),\dot{q}_i(t_i)$ of the system at a given time $t_0$, as well as the specific forces acting within the system and other quantities characterizing it, the observer can verify the correctness of the traced trajectories.
Moreover, the inertial observer can apply external forces $\vec{F}_e$ to the system $\mathfrak{S}$, as shown in Figure \ref{fig:io2}. We will refer to this observer as the interacting inertial observer (IIO).

By leveraging the concept of IIO, we now proceed to give a definition of the simulation concept, which, as we will see, also fits well with the type of simulation produced by virtual reality.

\begin{figure}[h]
\centering
\includegraphics[width=0.8\textwidth]{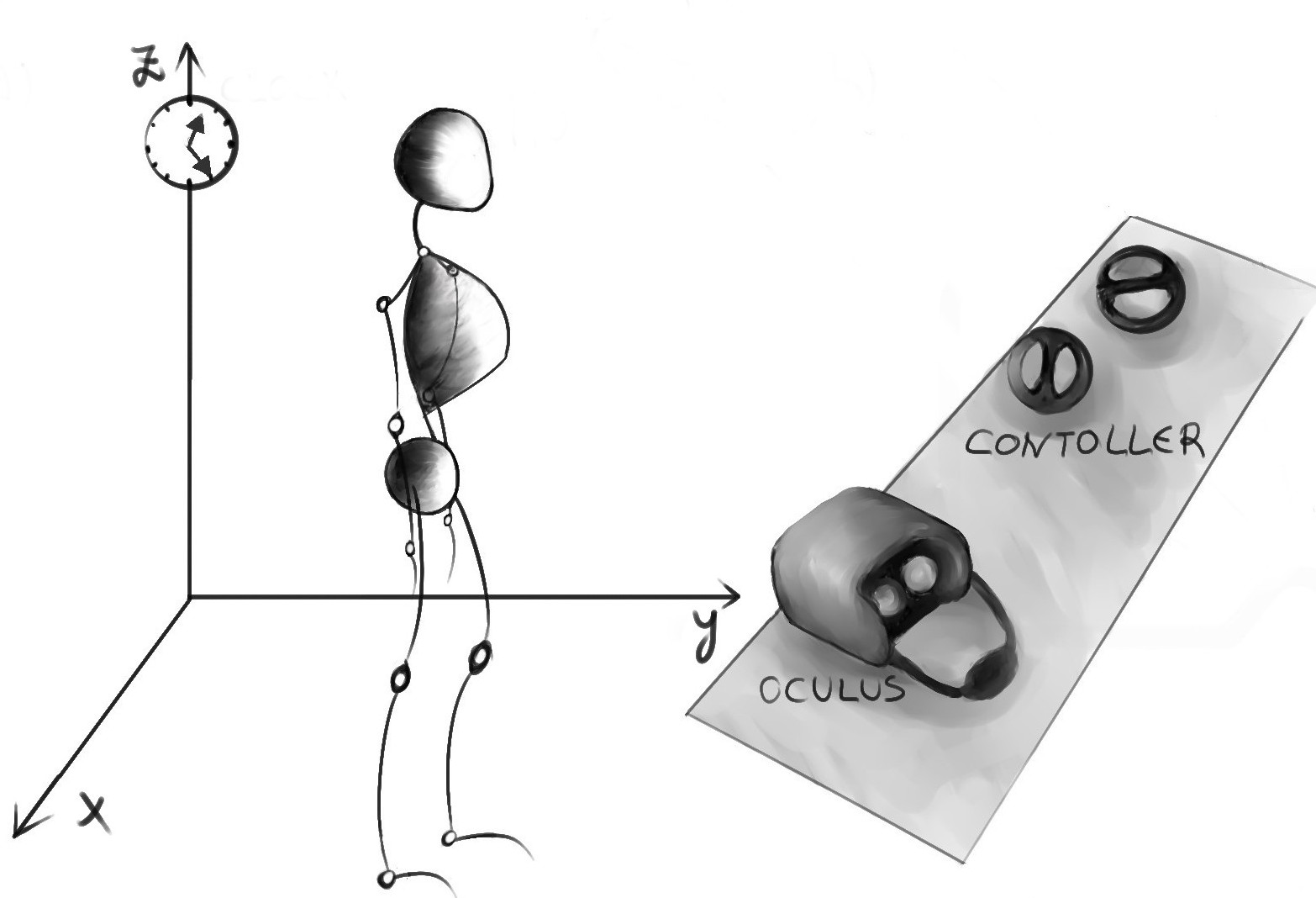}
\caption{The inertial observer is depicted alongside the basic equipment for immersing into virtual reality: Oculus and controllers.}
\label{fig:io3}
\end{figure}

\subsection{Simulation of a Physical System}
After introducing the concept of an observer and a physical system $\mathfrak{S}$, let's consider a physical system $\mathfrak{S}'$ described by $m$ variables $q'_i$ with $m\geq n$.
Suppose that for a certain time interval $\Delta t$ and certain initial conditions, the trajectory of $n$ of its coordinates corresponds to that of the coordinates of the system $\mathfrak{S}$. In such conditions, we say that the system $\mathfrak{S}'$ is an observable simulation of the system $\mathfrak{S}$.

The system $\mathfrak{S}'$ is observable in the same sense as $\mathfrak{S}$. From the observer's perspective, it exhibits $n$ observable coordinates that follow the same trajectory as the coordinates $q_i$. Therefore, an observer measuring its coordinates obtains the same results. In practice, the two systems could be indistinguishable to the observer.

In the case where $m$ is greater than $n$, we can assume that the observer decides to neglect the excess coordinates or hide them in such a way that they are disregarded.

By utilizing this concept, in the Results section, we will formulate a theorem that formalizes the relationship between a given reality $\mathfrak{S}$ and its simulation $\mathfrak{S}'$.

\subsection{Inertial Observer and Virtual Reality}
\begin{figure}
\centering
\includegraphics[width=0.8\textwidth]{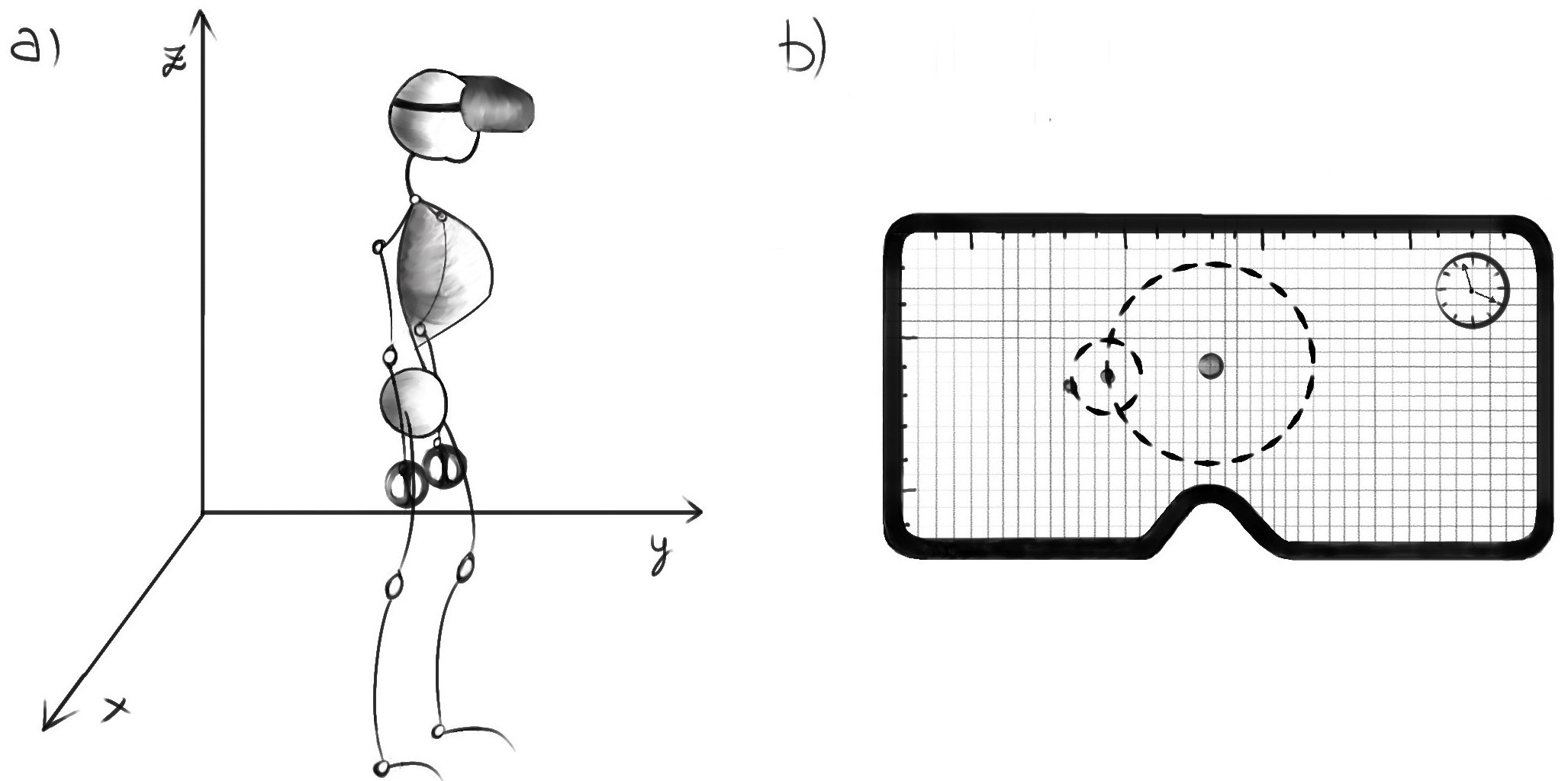}
\caption{The figure shows a) the inertial observer wearing Oculus and controllers, b) the view of the inertial observer generated by the Oculus.}
\label{fig:io4}
\end{figure}

In the previous section, we assumed that the IIO observes real physical systems. Now, let's consider that the IIO can wear equipment like the one described above (see Figure \ref{fig:io3}), consisting of an Oculus with a graduated scale and a clock, and a controller (see Figure \ref{fig:io4}).

When wearing the Oculus, the IIO observes the system $\mathfrak{S}''$ through the images produced by the pixels. The IIO can also interact with the system $\mathfrak{S}''$ using the controller.

In summary, we have seen that a given virtual reality, created to simulate a specific physical system $\mathfrak{S}$, is itself a physical system $\mathfrak{S}''$ consisting of the computer, software, and equipment.

\section{Results}

\newtheorem{teorema}{Theorem}

\begin{teorema}
Given a system $S$ and a set of initial conditions
\begin{equation}
\label{eq:condizioniiniziali}
q_1(0),...,q_n(0),\dot{q}_1(0),...,\dot{q}_n(0)
\end{equation}
with $n$ finite and greater than $1$, such that the trajectory of its representative point in the configuration space for a time interval $[0, 0+\Delta t]$ is the curve $C$, and that there exists a coordinate $q_j(t)$ on that curve that is continuous and monotonic within that time interval, then there exists a system $\mathfrak{S}'$ described by $m\geq n$ coordinates that is an observable simulation of $\mathfrak{S}$.\
Furthermore, there also exists a system $\mathfrak{S}'$ described by $n$ coordinates and $n-k$ constraints (with $1<k<n$) that is an observable simulation of $\mathfrak{S}$.
\end{teorema}

Before presenting the proof, let's analyze the meaning of the statement. The theorem asserts that for every finite system \cite{Deutsch}, there exists, or can be realized, a system that simulates it, thereby deceiving an observer whose observational capabilities are limited to measuring the system's coordinates. The second part of the statement further claims that there exists a system $\mathfrak{S}'$ with only $n-k$ degrees of freedom that simulates $\mathfrak{S}$, making it conceptually easier to handle.

\begin{proof}
We divide the proof into three cases, starting with the most obvious one: $m=n$, $m>n$, and $k<n$.

\paragraph{Case $m=n$:} The proof in this case is trivial since we can consider a system $\mathfrak{S}'$ such that $\mathfrak{S}' = \mathfrak{S}$, essentially an exact copy of the system, which is an observable simulation of the system itself.

\paragraph{Case $m>n$:} In this case as well, the proof is straightforward. We can consider $\mathfrak{S}' = \mathfrak{S}'_1 + \mathfrak{S}'_2$, where $\mathfrak{S}'_1 = \mathfrak{S}$ and $\mathfrak{S}'_2$ does not interact with $\mathfrak{S}'_1$.

\paragraph{Case $k<n$:} We limit the proof to the case $k=n-1$, as the other cases can be derived trivially from this one. The curve $C$ traversed by the representative point of the system $\mathfrak{S}$ can be expressed as a system of $n-1$ equations, each representing a hyperplane in the configuration space:

\begin{equation}
\label{eq:C}
\begin{cases}
f_1(q_1,...,q_n)=0 \\
f_2(q_1,...,q_n)=0 \\
\dots \\
f_{n-1}(q_1,...,q_n)=0\\
\end{cases}
\end{equation}

By considering the Jacobian of the system, from Dini's theorem, we obtain that the representation of the curve $C$ given in equation \ref{eq:C} is equivalent to the following:

\begin{equation}
\label{eq:vincoli}
\begin{cases}
q_1 = q_1(q_j) \\
\dots\\
q_j = q_j \\
\dots \\
q_n = q_n(q_j)\\
\end{cases}
\end{equation}
Considering that the equations \ref{eq:C} are formally identical to $n-1$ constraints expressed on the system $\mathfrak{S}$, then the equations \ref{eq:vincoli} represent the equations of motion for the $n$ coordinates $q_i()$ of the constrained system with the constraints defined by \ref{eq:C}.

Therefore, if there exists a $q_j$ that is continuous and monotonically increasing within the time interval $\Delta t$, the temporal evolution of the coordinates $q_i$ of the unconstrained system $\mathfrak{S}$ can be expressed as $q_i(t) = q_i(q_j)$, that is, within the time interval $\Delta t$ and under the initial conditions \ref{eq:condizioniiniziali}, the system $\mathfrak{S}'$ is an observable simulation of the system $\mathfrak{S}''$.
\end{proof}

\subsection{Example of Observable Simulation}

To provide an example of the concept of observable simulation, let's consider the planetary subsystem consisting of the Sun, Earth, and Moon, as shown in Figure \ref{fig:io1}. For simplicity, let's assume that the Sun remains in a fixed position unaffected by the Earth, and that the Moon is only influenced by the gravitational attraction of the Earth and not by that of the Sun. The coordinates of the system are therefore six, three for the Earth and three for the Moon, with no constraints.
\begin{figure}[h]
\centering
\includegraphics[width=0.8\textwidth]{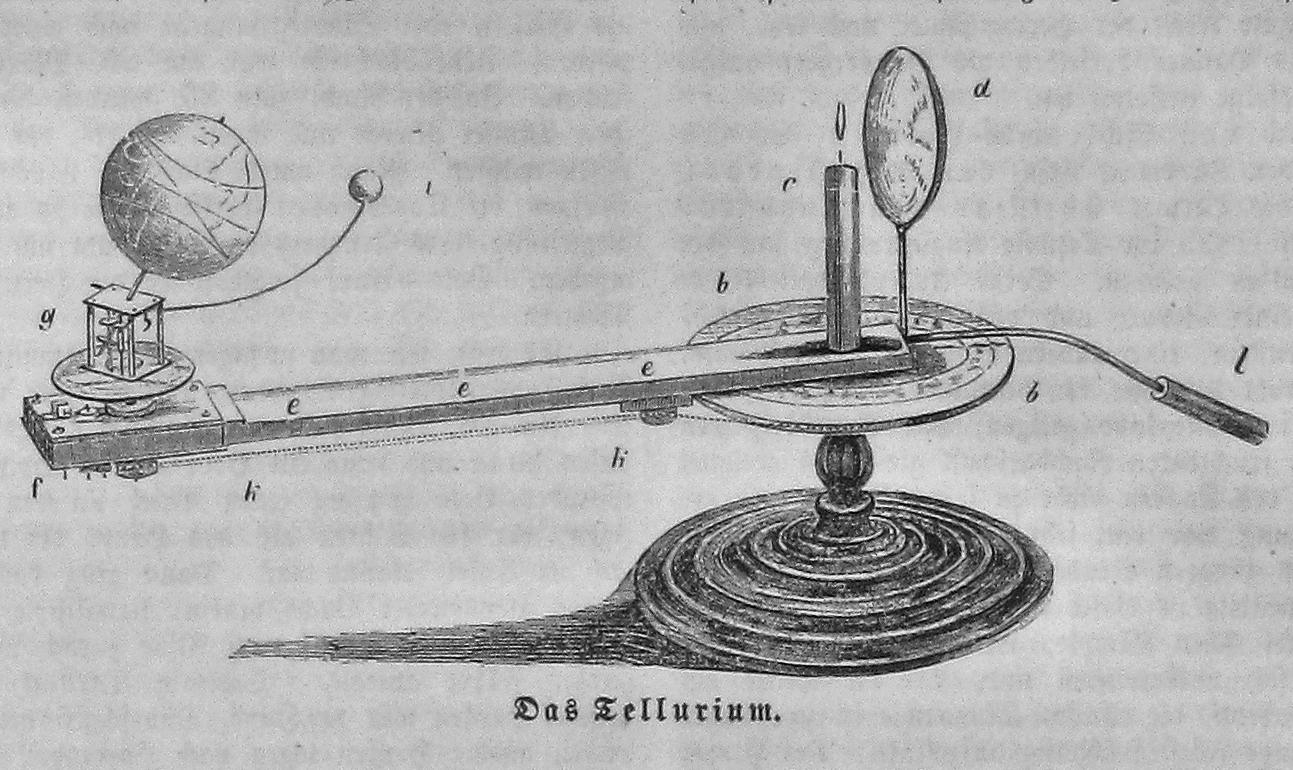}
\caption{The image shows a Tellurion, which is a simulator that reproduces the movement of the Sun, Earth, and Moon on a reduced scale. Used for educational and scientific purposes, the Tellurion provides a realistic visualization of the orbital and rotational motion of celestial bodies, offering a practical illustration of the relationships between the Sun, Earth, and Moon in the solar system. Image from Wikipedia.}
\label{fig:Tellurio}
\end{figure}

It is well known that under certain initial conditions, the orbits of the Earth and Moon approximate elliptical trajectories that can be approximated as circles. By choosing to represent the position of the Earth using a system of cylindrical coordinates ($\theta, \rho, z$), we can observe that the angle $\theta$ varies continuously and monotonically for any arbitrarily chosen interval $\Delta t$. Therefore, based on the previous proof, it is possible to realize a system $\mathfrak{S}'$ that is an observable simulation of $\mathfrak{S}$. The Tellurion, shown in Figure \ref{fig:Tellurio}, is a concrete, scaled implementation of the system $\mathfrak{S}'$.

\begin{figure}[h]
\centering
\includegraphics[width=0.8\textwidth]{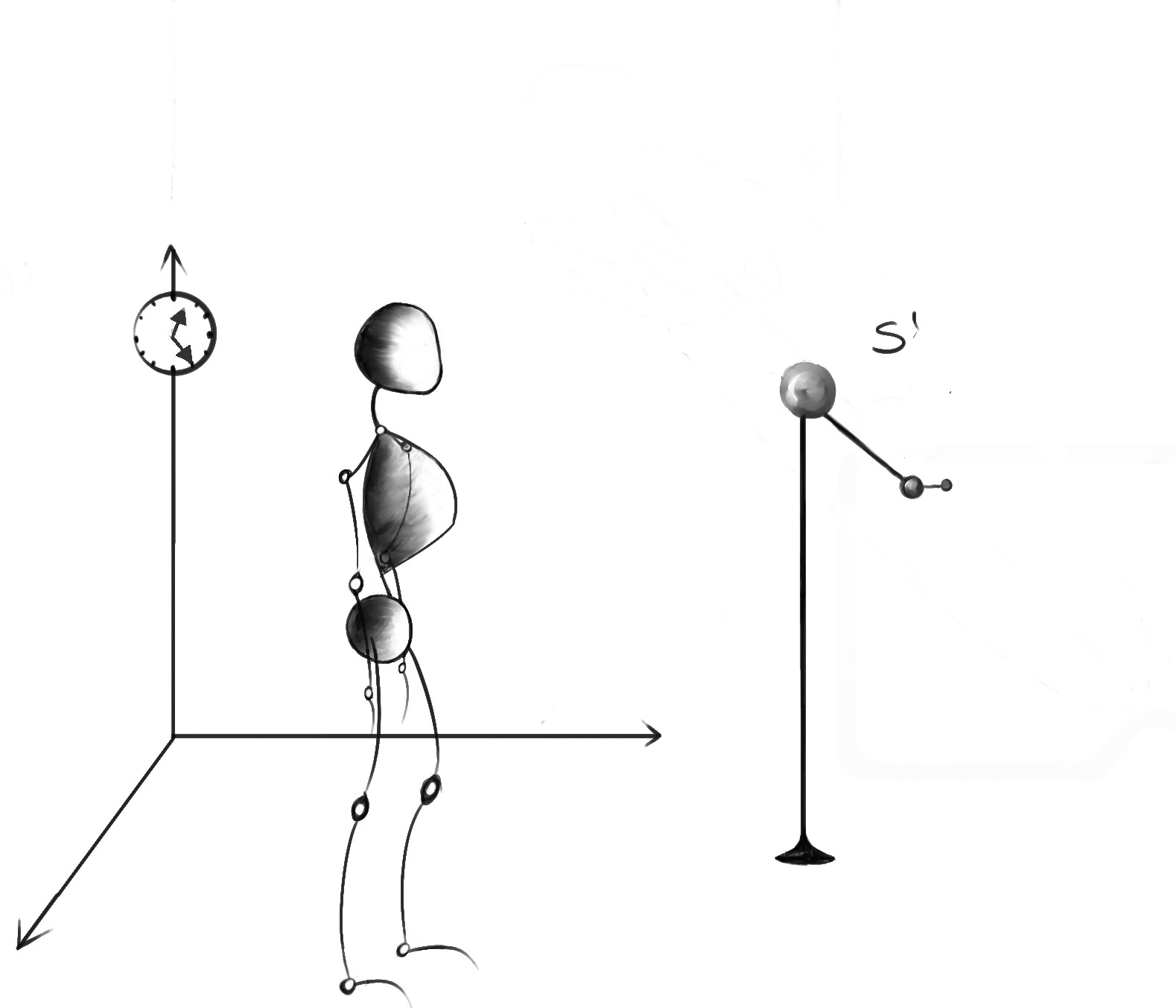}
\caption{The image shows the observer and the system $\mathfrak{S}'$, which is a Tellurium simulating a specific orbital condition of the Sun-Earth-Moon system.}
\label{fig:io2b}
\end{figure}

It is a mechanism with a single degree of freedom that, using a set of gears similar to those in a mechanical clock, produces the two circular movements of the Earth and Moon while respecting the relationship between the orbital periods of the two celestial bodies. The Tellurion allows the observation of the three spatial coordinates of the Earth and the Moon. It serves as an observable simulation (to scale) of the Sun-Earth-Moon system. Figure \ref{fig:io2b} depicts the IIO observing the system $\mathfrak{S}'$ represented by the Tellurion.

\begin{figure}
\centering
\includegraphics[width=0.8\textwidth]{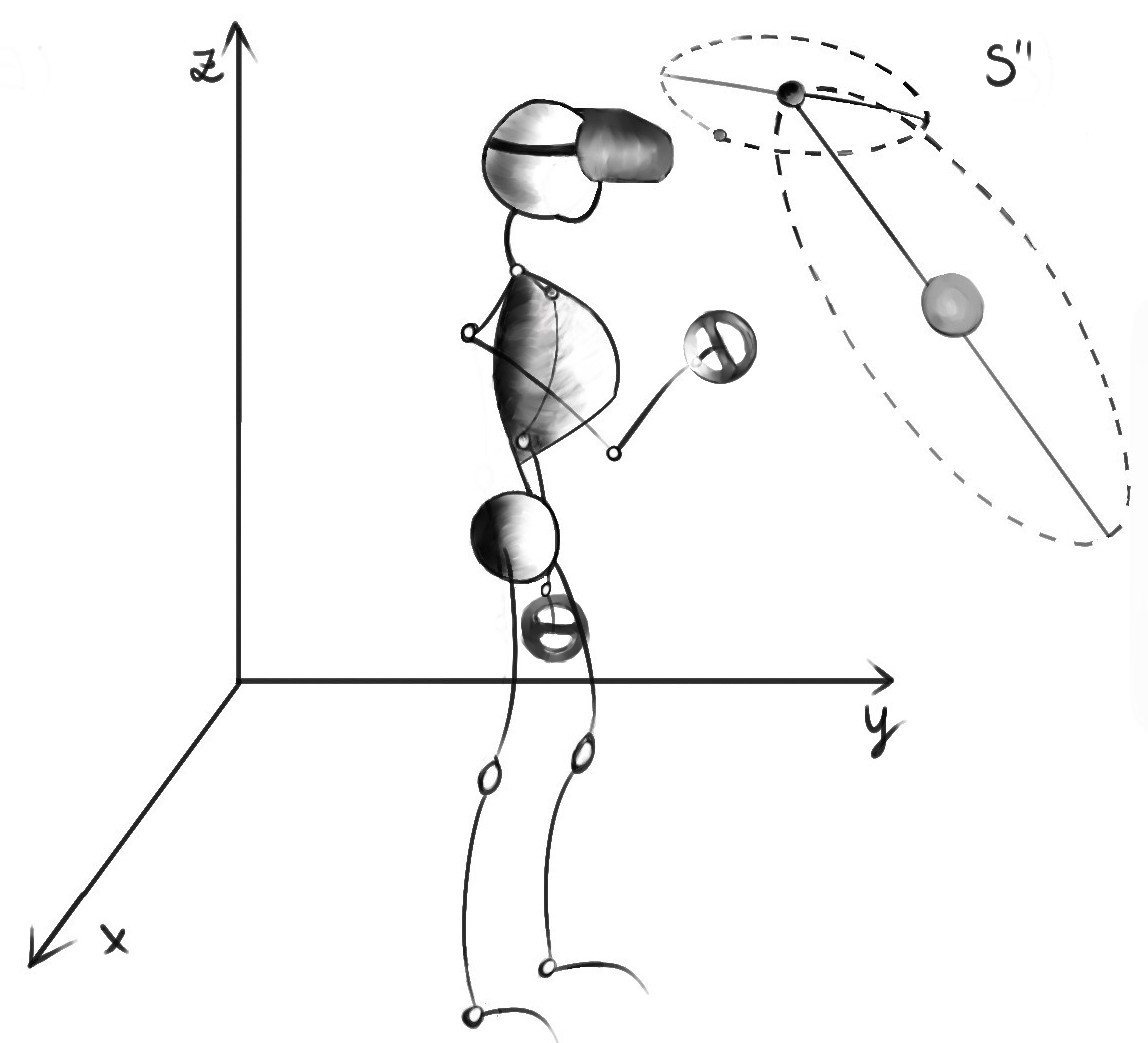}
\caption{The inertial observer wears Oculus and controllers and can measure the position of celestial bodies produced by the virtual reality system $\mathfrak{S}''$.}
\label{fig:io5}
\end{figure}

Now let's consider that the IIO, after observing the system $\mathfrak{S}$ as described in Figure \ref{fig:io1}, wears the equipment shown in Figure \ref{fig:io3}, which virtualizes the planetary system (Sun-Earth-Moon) as depicted in Figure \ref{fig:io4}. Although aware of wearing an Oculus, the IIO, by measuring the position and time of what is observed, formally concludes that they are observing the planetary system, as shown in Figure \ref{fig:io5}. Since virtual reality is still a physical system, it also formally represents a simulation that we will call $\mathfrak{S}''$ of $\mathfrak{S}$. In terms of software, we can see that, neglecting temporary variables required for computational purposes (indices, flags, etc.), the number of variables to describe the simulation is only one, as in the case of the Tellurion.

\section{Discussion}

In this section, we will attempt to answer the questions posed in the introduction. Let's assume that the IIO observer is observing a system $\mathfrak{S}'$ that is a copy of a system $\mathfrak{S}$ with the same coordinates and constraints. Under these conditions, the observer is clearly unable to distinguish between $\mathfrak{S}$ and $\mathfrak{S}'$. This simple observation partially answers the question at hand because it is always possible to induce the observer to agree that what they are measuring is either the system $\mathfrak{S}$ or an identical system. This result is obvious since, in the assumptions about the system $\mathfrak{S}'$, we are stating that it coincides with the system $\mathfrak{S}''$ in every measurable aspect. Therefore, the observer cannot measure any differences between the two.

\begin{figure}
\centering
\includegraphics[width=0.8\textwidth]{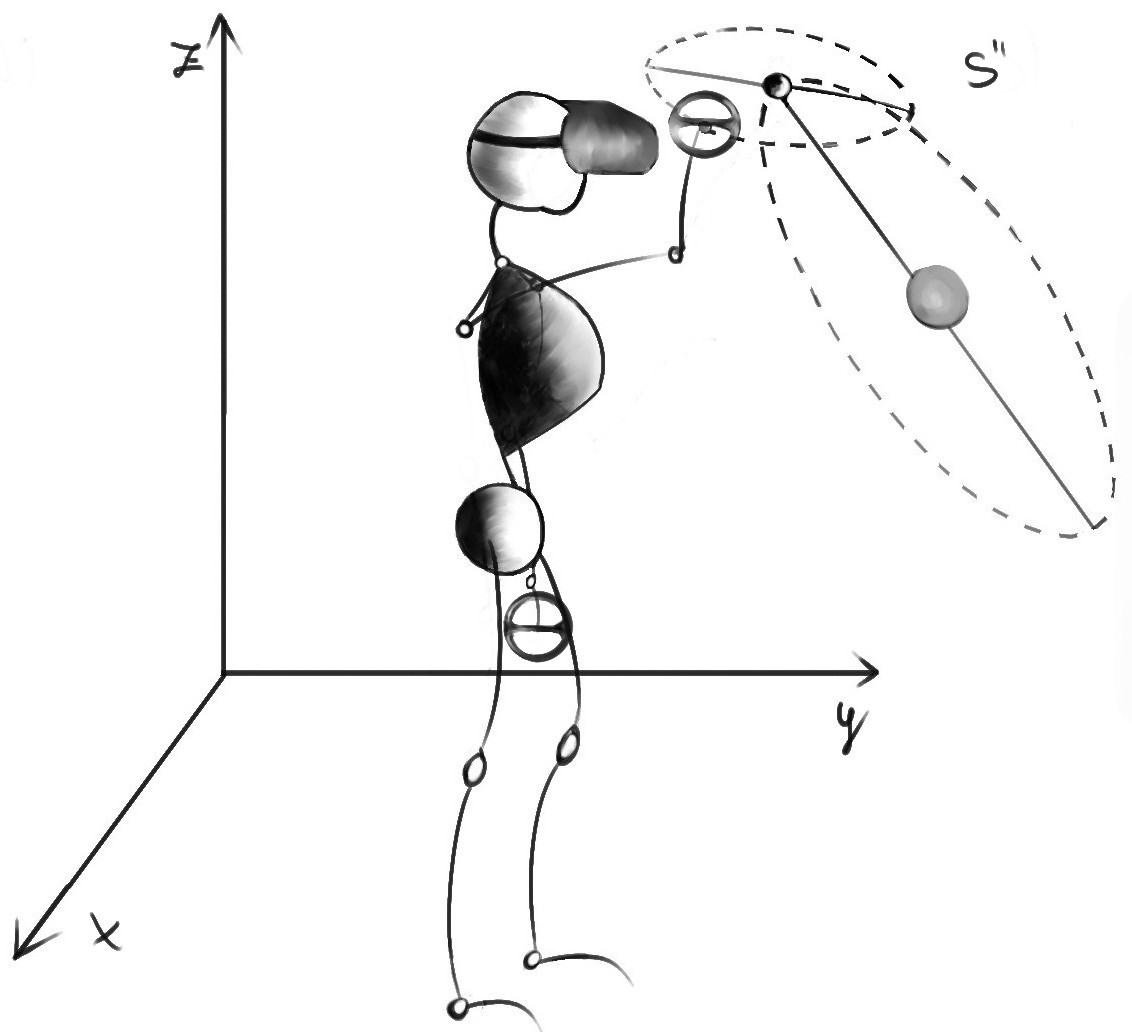}
\caption{The inertial observer applies a force $\vec{F}_e$ to the physical system $\mathfrak{S}$.}
\label{fig:io6}
\end{figure}

Practically, this tells us that if virtual reality consisted of creating an environment that coincides with the reality we want to simulate, the result would always be certain and positive. However, this type of simulation does not offer any advantages over reality. For example, if one wanted to train a pilot under risky flying conditions by placing them in such conditions, it would only risk the pilot's life and the cost of the vehicle. Similarly, it would (in principle) be possible to study lunar phases by creating an exact replica of the Sun-Earth-Moon system.\\
We have also demonstrated that it is possible to create an observable simulation using fewer degrees of freedom than those of the system $\mathfrak{S}$. In the example we proposed, we saw that the Tellurion is a system $\mathfrak{S}'$ described by six coordinates, and it has only one degree of freedom, which can be defined as an observable simulation of the system $\mathfrak{S}$ consisting of the Sun-Earth-Moon. Consider the unrealistic hypothesis of creating a Tellurion that respects the real distances and dimensions and asking the observer to distinguish it from the system $\mathfrak{S}''$. Equipped with a clock and a meter, the observer will not be able to recognize any differences. In this regard, consider that the observer's meter has a certain resolution, so to ensure that they cannot distinguish between the two systems, it is necessary and sufficient to create $\mathfrak{S}'$ with an appropriate resolution and eventually vary it according to the resolution of the observer's meter.\\
However, the two proposed simulation systems differ significantly in the number of degrees of freedom. So, we wonder what happens if the observer decides to interact with the degrees of freedom of the system? Let's imagine that the observer can apply an arbitrary force, in terms of magnitude, direction, and orientation, to any coordinate of the system. If the observer is observing a simulation of the Moon's motion based on a replica of the system, they can apply a force to the simulated Moon that alters its orbit.
In this case, they will measure the change in trajectory and find that the new trajectory conforms to the law of universal gravitation, and therefore, they will maintain the belief that the observed system is part of the solar system or indistinguishable from it. On the other hand, if they are observing the Tellurion (at a $1:1$ scale), they cannot apply a force to the object representing the Moon without also applying it to the entire Tellurion because the only degree of freedom is the angle $\theta$.
In practice, they will either freeze the system or modify the overall velocity of the celestial bodies. In any case, the trajectory of the Moon after applying the force will be incompatible with the central force gravitational model that they believe they are observing. Thus, in this situation, the IIO would be able to distinguish the simulation from reality.\\
The same reasoning applies to virtual reality. The IIO observing the system $\mathfrak{S}''$ and attempting to apply a force as shown in Figure \ref{fig:io6} will see the laws of physics being violated because the only degree of freedom used by virtual reality to simulate the system is not sufficient to simulate the correct trajectory after the external force $\vec{F}_e$ is applied by the IIO.

Therefore, we can distinguish between two types of simulations: those that are observable and interactive, and those that are only observable but not interactive.

In practice, we see that if we want a system $\mathfrak{S}'$ to be an interactive simulation of $\mathfrak{S}$, it must have an equal or greater number of degrees of freedom than the system $\mathfrak{S}$. Otherwise, it is possible to create an observable simulation $\mathfrak{S}'$ of the system $\mathfrak{S}$, but it can be distinguished from $\mathfrak{S}$ by properly interacting with the system.

After these premises, let's finally address the questions posed in the introduction and see how the argument developed so far relates to virtual and immersive reality in the commonly used sense.

The computer-software-equipment trio is a physical system with $m$ coordinates and $p$ degrees of freedom. As we have seen, if $p$ is less than the number of degrees of freedom $n$ of $\mathfrak{S}$, then virtual reality will only simulate $\mathfrak{S}$ in terms of observability but will not be a fully interactive reality. In other words, there will be actions that the observer can take that will allow them to formally distinguish what is simulated from the system $\mathfrak{S}$ that virtual reality claims to simulate. If $p$ is greater than or equal to $n$, virtual reality could simulate reality in an interactive manner and become \textit{indistinguishable} from it based on the terms adopted so far. Therefore, we can answer the first question as follows:

\begin{itemize}
\item It is possible to induce a (rational) observer to agree that reality and simulation are indistinguishable if reality is simulated using an equal or greater number of degrees of freedom than the reality itself.
\end{itemize}

As for the second question, the answer is obviously negative. Since the computer-software-equipment trio is a system subject to the laws of physics, any result produced by it can always be represented by the observer as a representative point of a system in the configuration space. However complex its trajectory may be, it will always be describable by time laws compatible with the laws of physics. Therefore, we can state:

\begin{itemize}
\item It is not possible to simulate a reality where the laws of physics do not hold.
\end{itemize}

\section{Conclusion}

The simulation of classical and finite realities is always possible, but if one wants to completely "trick" the user and make them unable to distinguish reality from its virtualization, it is necessary to employ at least the same number of degrees of freedom that describe the real system. Considering that our objective reality is made up of particles (atoms, molecules, etc.) that in principle can be measured by the user, this result sets a limit on what can objectively be virtualized.

In conclusion, virtual reality can create highly immersive and realistic simulations, but there will always be ways for a perceptive and interactive observer to discern between the simulated reality and the actual reality based on the number of degrees of freedom available for interaction. However, for practical purposes and within the limits of human perception, virtual reality can provide experiences that are effectively indistinguishable from reality.

\end{document}